*Article*

# Growth Dynamics of *S. aureus* with Sugars and Sugar Alcohols in Weak Magnetic Fields

**Samina Masood*, Angel Arrieta and Derek Smith**

Department of Mathematical and Physical Sciences, University of Houston Clear Lake
* Correspondence: masood@uhcl.edu

**Abstract**

We study the effects of weak magnetic fields (around 2 mT) on the growth of *Staphylococcus aureus* (*S. aureus*) in the presence of a few sweeteners (monosaccharides, disaccharides, sugar alcohols, and consumer-grade sweeteners). Bacterial growth rates were compared in various magnetic fields at room temperature. Bacterial growth was estimated using optical absorbance measurements at various wavelengths, and pH values were manually estimated using pH strips. Absorbance was measured at 492 nm and 630 nm, which are wavelengths comparable to the size of a cell of *S. aureus* after division. This comparability plays a vital role in the scale of measured absorbance values. The results imply that bacterial growth may be reduced due to acidic byproducts formed by metabolizing sugars or sugar alcohols, as an increasingly acidic solution is less ideal for bacterial growth. Magnetic fields were also found to have a minor effect on pH estimates. These results reveal potential effects on microorganisms in the presence of sugars and sugar alcohols in addition to weak magnetic fields, demonstrating the contribution of various environmental conditions with increasing prevalence in the modern day.

## 1. Introduction

Bacterial cells divide using binary fission. There are a great variety of bacterial species with various strains, and their short lifetimes can easily undergo mutation depending on the environmental conditions and available nutrients. A typical mass of bacterium can be considered as 1 pg (picogram) or $1\times10^{-12}$ grams, which is $10^{11}$ times heavier than a single atom (1 amu = $1.66 \times 10^{-24}$ g) [1]. It can be clearly seen that the rate of flow of bacterium cells is negligible as compared to nutrients (ions) in the bacterial culture, especially in response to weak external forces on the fluid including weak magnetic fields which direct the charged ions through the magnetic field lines. Nutrients in a solution move faster than non-motile bacteria due to their smaller size and mass, and the relative change in the mobility of ions depends on their relative charges and masses.

In the modern world, there are various sugars and sugar alcohols that are commonly consumed, and we are frequently near electronics that produce weak magnetic fields. We therefore take an interest in understanding how sugars, sugar alcohols, and weak magnetic fields (in the order of 1 mT) together might affect bacterial cells' growth and properties.

The most commonly known pathogenic bacterium, *E. coli*, is gram-negative, rod-shaped, and has been extensively studied. We instead focus on *Staphylococcus aureus* (*S. aureus*) in this paper. It is a round-shaped gram-positive bacterium and is commonly found in human bodies. However, its infections are not uncommon and are difficult to treat as it can develop resistance against antibiotics [2].

In this paper, the "sweeteners" we chose to study include the sugars fructose, glucose, lactose, maltose, and sucrose, as well as the sugar alcohols glycerol and sorbitol. Additionally, we have studied consumer-grade table sugar and blue agave syrup as they are examples of common, accessible sweeteners which are frequently consumed.

Fructose and glucose are monosaccharides ($C_6H_{12}O_6$) commonly found in plants such as fruits or vegetables, whereas lactose is a disaccharide ($C_{12}H_{22}O_{11}$) formed by combining galactose and glucose naturally found in dairy products. Maltose is a disaccharide of two glucose molecules that can be produced by malting grain, and sucrose (or table sugar) is a disaccharide composed of fructose and glucose.

Glycerol ($C_3H_8O_3$) and sorbitol ($C_6H_{14}O_6$) are sugar alcohols commonly used in drugs and in some sugar-free products as sugar substitutes. Glycerol can be sourced from animals and plants, while sorbitol is commonly found in fruits and vegetables.

We have chosen to test consumer-grade table sugar and blue agave syrup (fructose, glucose, and water) as additional experimental variables because of their accessibility and impurities. These consumer-grade sweeteners are more likely to represent the quality of product consumed by an average person compared to lab-grade chemicals, though the differences may be negligible for our tests. Blue agave syrup is made from the plant Agave *tequilana,* and the composition of the syrup is largely fructose, followed by glucose, water, and trace amounts of sucrose [3], although the exact composition depends on the manufacturer.

Previous studies show that various bacterial species such as *E. coli*, *P. aeruginosa*, *B. subtilis*, *S. aureus*, and *S. epidermidis* are affected differently in various weak magnetic fields [4-5] or if they are grown on various surfaces [6-7]. In a prior study, we tested *S. aureus* and its interaction with glycerol and lactose in the presence of various weak magnetic fields at room temperature and compared with our previous results. Bacterial growth was estimated using optical absorbance measurements at various wavelengths, and pH strips were used to estimate pH values of the samples. It has also been noticed that a constant magnetic field may cause a kind of swirling effect and that bacteria tend to gather along the magnetic field lines. This swirling effect is sometimes indicated in previously published images, but they were not highlighted as the images did not show clear patterns. *E. coli* is known to have the property to aggregate as a stress response and to create biofilms [8].

*S. aureus* had been known for its ignorable motility. However, it was noticed in 2015 that *S. aureus* can move across an agar surface [9]. The spreading motility was later observed using a modified assay which showed a branching structure emerged from the central colony. It was also noticed that the bacteria can spread like a comet near the surface. In 2017, it was observed by the same group that bacterial movement is critical for its survival in a diverse environment [10].

Existing results from prior experiments also provide motivation for the study of bacterial motion in the presence of weak magnetic fields. It has been noticed that various weak magnetic fields affect the movement of bacteria differently. For *S. aureus*, previous results have found that the bacteria adapt to static, uniform magnetic fields more readily than other magnetic field types, resulting in an increased growth rate in the magnetic field. We believe weak magnetic fields may perturbatively affect motility and therefore the growth rate of various bacterial species according to their size and shape. The movement and availability of nutrients in growth media within a magnetic field also depends on the mass and charge of the nutrients and may affect bacterial growth. We determine optical density to estimate the growth rate [11]. Usually, bacterial concentration is measured using 580 nm or 600 nm whereas sometimes larger wavelengths are used for this purpose [12]. The measuring wavelength is highly dependent on the bacteria size and nutrient broth solution since it is important to maximize refraction by the target bacteria while minimizing interaction from other elements such as the broth and nutrients [13-14]. Due to this, it is important to consider changes in bacteria size due to variables such as sweeteners or magnetic fields that could affect said optimal wavelength. It is therefore important to compare different measuring wavelengths.

We have held a preliminary study about the growth of bacteria in nutrient broth containing various sugars and sugar alcohols such as glucose, glycerol, fructose, lactose, and sucrose. Because glycerol is not chemically a carbohydrate, we will group these sugars and sugar alcohols together as "sweeteners" for simplicity as an alternative to more complicated terms such as "polyhydroxy compounds". In this study, we found that the change in pH value of samples containing glycerol ($C_3H_8O_3$) was minimal compared to samples grown without sweeteners, but the change in pH of samples containing lactose ($C_{12}H_{22}O_{11}$) was still noticeable. Therefore, these sweeteners are interesting subjects for testing bacterial growth in the presence of sweeteners. Moreover, glycerol is a sugar alcohol that is used in several sugar-free products and pharmaceutical products. Its detailed study therefore attracts particular interest from researchers. On the other hand, lactose is particularly important because it is found in many dairy products and a study has shown that lactose inhibits bacterial growth [15].

The pH value is a measure of the alkaline or acidic properties of a material. Human bodies can have pH values between 7.35 and 7.45 within blood and interstitial fluid [16], which is slightly alkaline as it is higher than 7 (neutral). The pH value is an extremely sensitive parameter, as a minor change in these values can cause severe consequences in human bodies [16]. It is therefore particularly important to know how the pH value is affected by human-incubated bacteria in the presence of various sweeteners. Therefore, the measurement of pH value is very important and attracts the interest of researchers.

The next section describes the experimental results and highlights the findings. The Materials and Methods section is included afterward to describe the experimental procedure in detail.

## 2. Results and Discussions

The bacterial samples grown in microplates showed similar patterns over time for most wells. However, some of them showed a decline much earlier than others. This may be due to a minute difference in quantity of broth or inoculation as these wells contain very small amounts. On the other hand, outliers frequently appeared in the $12^{th}$ well (the rightmost well) of each row. Therefore, to reduce the impact of this seemingly position-dependent error, the twelfth samples for each sweetener was ignored.

Figure 1 shows the average absorbance for each sweetener solution. Each graph shows the general trend of more growth observed with the shorter wavelength (492 nm) as compared to the longer wavelength (630 nm). Given the size dependence of bacteria detection based on wavelength [14], this could be an indication of a greater concentration of small-sized bacterial cells as compared to larger ones, considering that smaller bacteria can skip detection with larger wavelength.

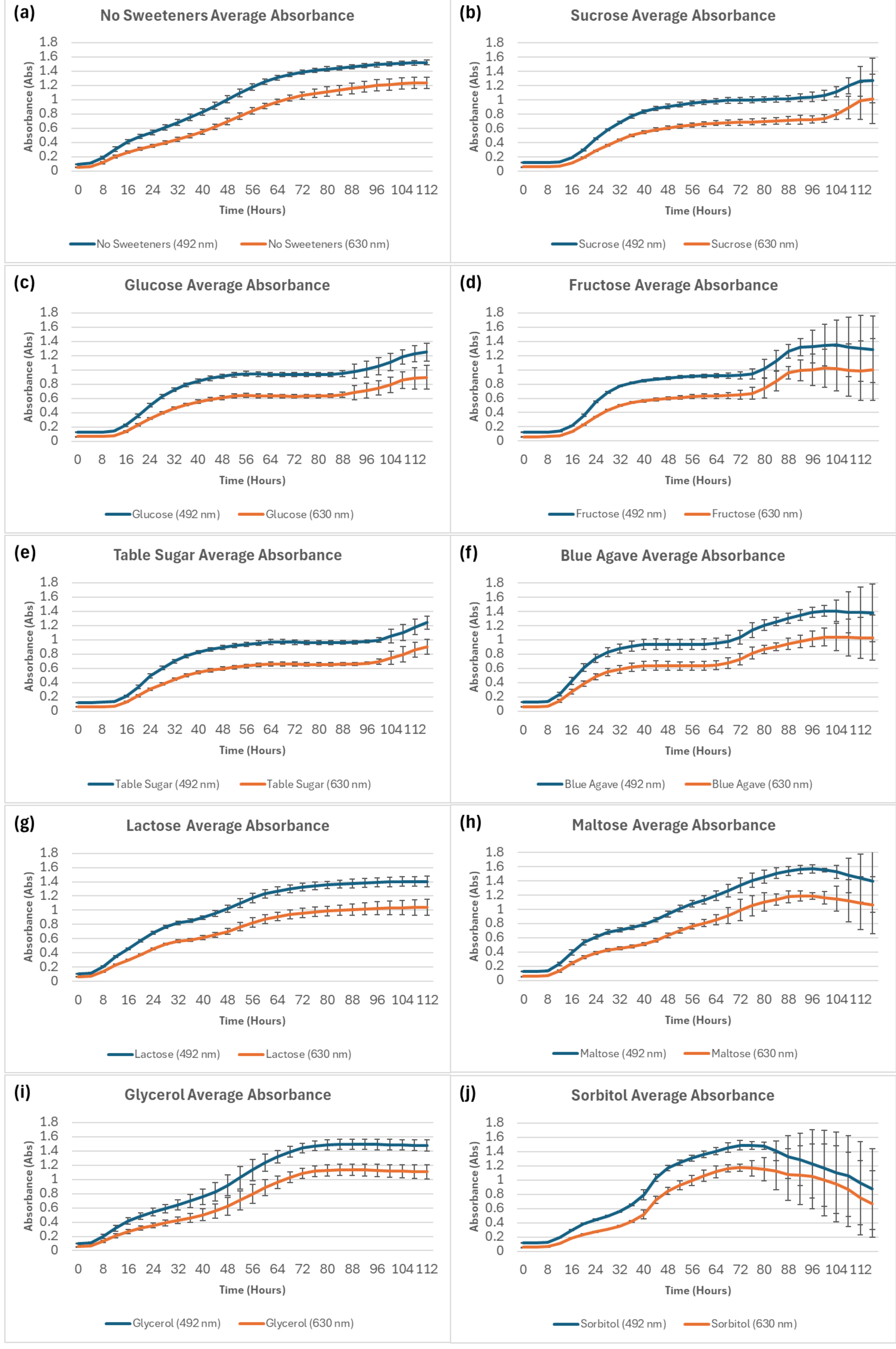

**Figure 1.** Average **a**bsorbance for wells containing *S. aureus* samples. Each graph displays data obtained using 492 nm and 630 nm wavelength light sources**.** The wells contained LB broth solutions with *S. aureus* bacteria along with one type of sweetener. (a) No Sweetener, (b) Sucrose, (c) Glucose, (d) Fructose, (e) Table Sugar, (f) Blue Agave, (g) Lactose, (h) Maltose, (i) Glycerol, or (j) Sorbitol. 11 wells were used for each sweetener, and measurements were taken for 112 hours in 4-hour intervals.

Graphs of the average absorbance in Figure 1 show three general growth pattern cases for the provided types of broth. In the case of sucrose, fructose, glucose, table sugar, and blue agave, growth briefly stops after 32 hours, after which growth resumes once again but at differing times for each sugar. For the second case appearing with no sweeteners, lactose, maltose, and glycerol, they experience continuous growth which stops at a stationary phase reached at times after 72 hours. Lastly for the third case with sorbitol and maltose, continuous growth is experienced for about 72 hours but then followed by a death phase without a stationary phase.

Because table sugar can be considered the consumer-grade version of sucrose and blue agave syrup is mostly comprised of fructose with some glucose and trace amounts of sucrose, it is expected that sucrose, fructose, glucose, table sugar, and blue agave should have similar growth patterns due to their chemical compositions. That said, considering that maltose has a different growth pattern, it is possible that, although it is composed of two glucose molecules, its disaccharide configuration could play a role on growth dynamics, similarly, lactose also presents a differing pattern even though it is also made of glucose, however, in this case, the presence of galactose could result in differing growth patterns.

Both glycerol and maltose show a slight standard deviation during the growth phase, but otherwise, the highest standard deviations are present during late stages of growth for most samples. Two exceptions include the unsweetened samples, which show minimal deviation throughout the growth period, and glycerol, which shows its largest deviation halfway through the 5-day growth period and diminishes towards the end.

The standard deviation of the samples was greater in the samples with sorbitol and tended to be the smallest in the samples without sweeteners. This trend is consistent for measurements at both 492 nm and 630 nm. The high standard deviation shown with sorbitol is due to several plate reader cells experiencing decline early shown in Figure 2. That said, all sample cells eventually showed a decline by the end of the 5-day period.

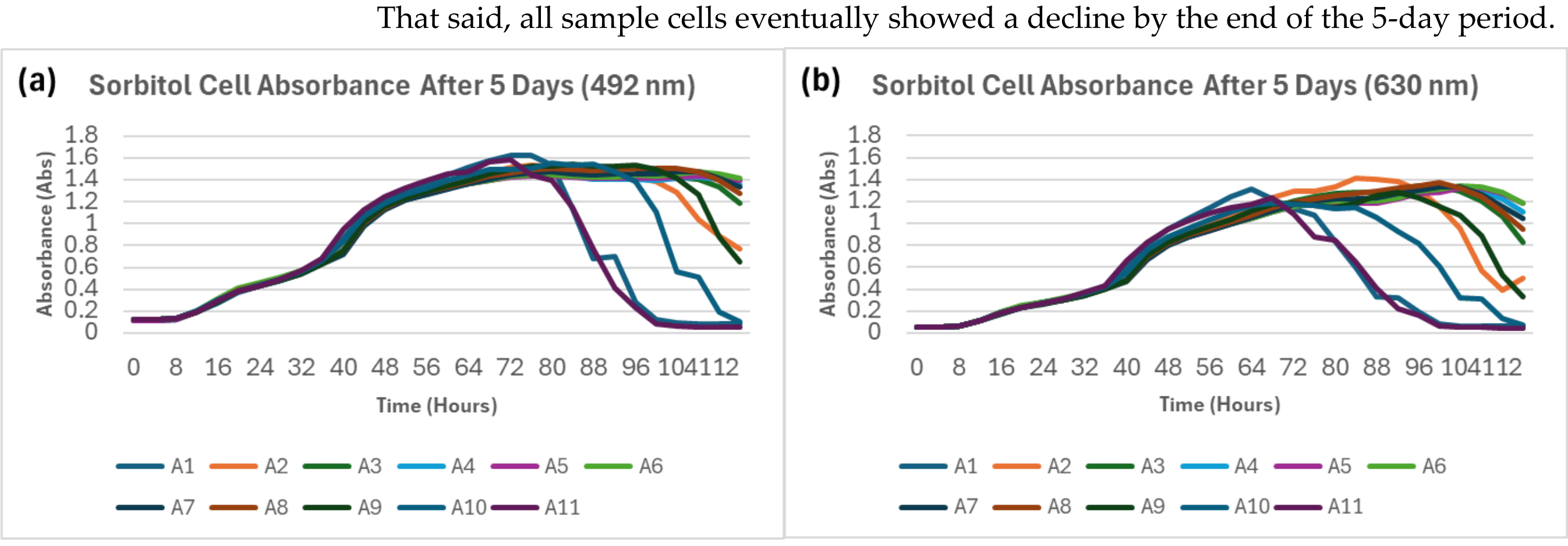

**Figure 2.** Individual cell absorbance for Sorbitol samples for (a) 492 nm and (b) 630 nm wavelengths. Sorbitol samples were grown in row A of the microwell plate from which 11 samples are displayed.

From the absorbance data, the largest growth for bacterial samples occurred in samples with maltose, reaching the highest absorbance with an average value of 1.569 (for 492 nm) at 92 hours with a standard deviation of 0.057. Comparing all samples, however, samples without any sweetener show the highest growth from beginning to end of the 5-day period. This data is shown in the following table, where the difference in growth can be easily noticed.

**Table 1.** Comparative growth of various bacterial samples with or without sweeteners.

| Average Absorbance | Initial Value | Final Value | Growth Ratio |
|---|---|---|---|
| No Sweeteners, 492 nm | 0.098909091 | 1.522518182 | 15.39310662 |
| No Sweeteners, 630 nm | 0.054954545 | 1.240809091 | 22.57882548 |
| Fructose, 492 nm | 0.123392 | 1.187842 | 9.626572225 |
| Fructose, 630 nm | 0.062242 | 0.926092 | 14.87889207 |
| Glucose, 492 nm | 0.123958 | 1.160467 | 9.361775763 |
| Glucose, 630 nm | 0.065208 | 0.833525 | 12.78255735 |
| Lactose, 492 nm | 0.109416667 | 1.413883333 | 12.92201066 |
| Lactose, 630 nm | 0.064583333 | 1.043541667 | 16.15806452 |
| Maltose, 492 nm | 0.12875 | 1.288142 | 10.00498641 |
| Maltose, 630 nm | 0.059767 | 0.976767 | 16.34291499 |
| Sucrose, 492 nm | 0.120825 | 1.172933 | 9.707701221 |
| Sucrose, 630 nm | 0.059633 | 0.935133 | 15.68146831 |
| Glycerol, 492 nm | 0.101183333 | 1.483708333 | 14.66356449 |
| Glycerol, 630 nm | 0.055608333 | 1.102608333 | 19.82811329 |
| Sorbitol, 492 nm | 0.120217 | 0.8061 | 6.705374448 |
| Sorbitol, 630 nm | 0.055825 | 0.61245 | 10.97089118 |
| Blue Agave, 492 nm | 0.129942 | 1.269775 | 9.771859753 |
| Blue Agave, 630 nm | 0.06345 | 0.952508 | 15.01194641 |
| Table Sugar, 492 nm | 0.1217 | 1.157042 | 9.507329499 |
| Table Sugar, 630 nm | 0.05985 | 0.841333 | 14.05736007 |

The growth ratio was calculated as the ratio between the initial and final average values as shown in Table 1. This allows us to estimate the relative cellular growth between samples. However, the relationship between absorbance and cell concentration is not necessarily linear [14]. Testing with different microplates, plate readers, or wavelengths could affect the results due to minor, compounding differences in accuracy and optical properties. Furthermore, error margins become more significant for smaller values. It is also worth mentioning that absorbance is not equivalent with optical density, although both are often used interchangeably. However, it gives a good estimate of optical density. Absorbance measures the light absorbed at a given wavelength, while optical density includes both the absorption and scattering of light. Optical density is the term commonly used for optical estimations of cell concentrations.

Typical *S. aureus* size is within the range of 0.5 to 1.5 micrometers, which corresponds with 500 to 1500 nanometers. This is the reason why the wavelength under 500 nanometers can detect more bacteria (larger optical density), as these waves are smaller than the expected size of bacterial cells. However, the size of bacteria cells may vary depending on the environment and the availability of nutrients.

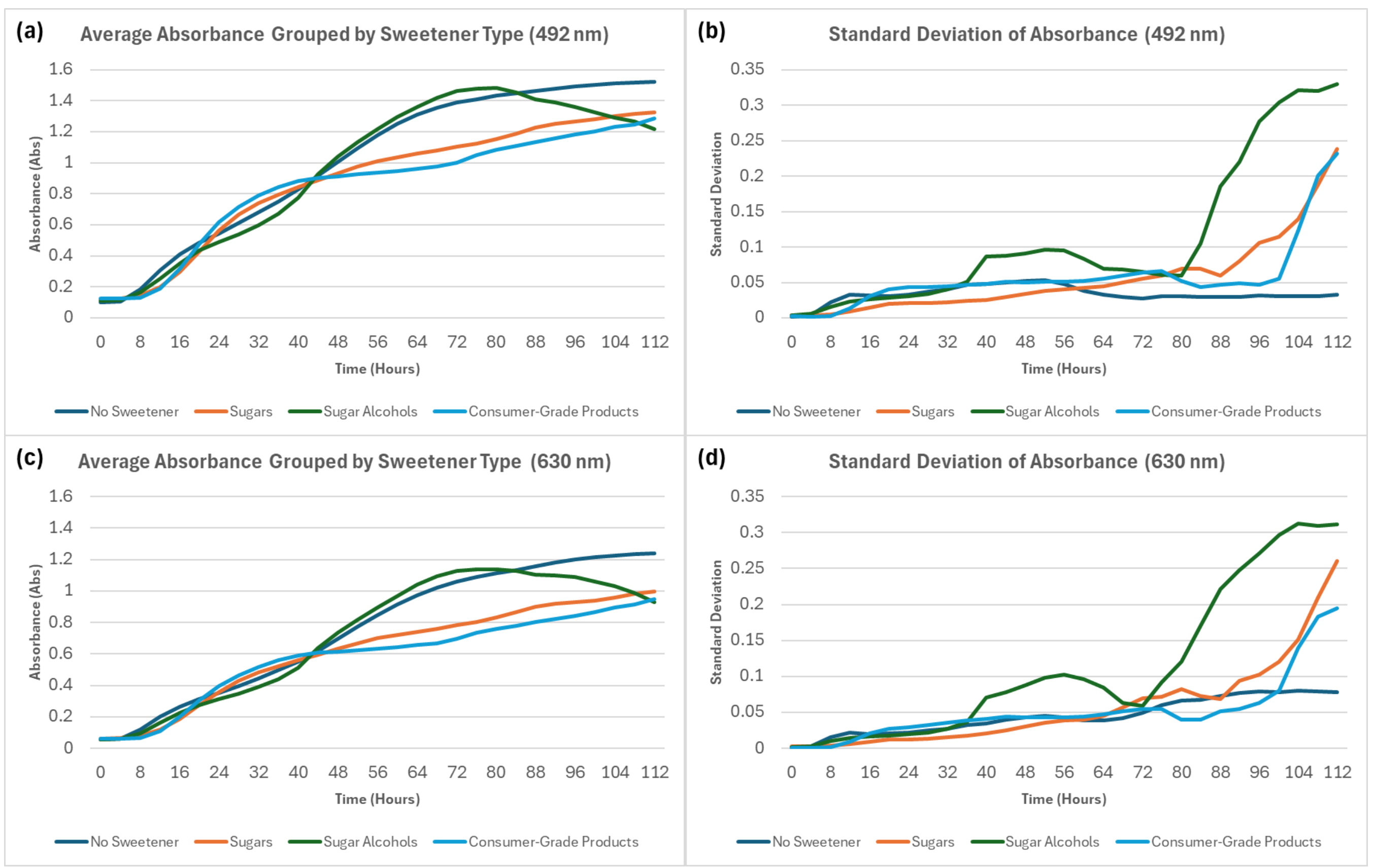

**Figure 3.** Averages and standard deviations of absorbance measurements for each sweetener type, separated by wavelength. (a) Average of absorbance values of "sugars" at 492 nm, (b) average of standard deviation of values in the "sugars" at 492 nm, (c) average of absorbance values of "sugars" at 630 nm, (d) average of standard deviation of values in the "sugars" at 630 nm.

Graphs of the average absorbance per sweetener type in Figure 3 show an initial faster growth for both lab-grade and consumer-grade sugars, with a slower growth for samples containing either no sweeteners or containing sugar alcohols. However, this relationship inverts at the 44-hour mark where the unsweetened and sugar alcohol samples show a faster growth instead (for the provided amount of broth). The standard deviation also tends to increase throughout the period, but it maximizes for sugar alcohol samples after approximately 100 hours, which corresponds to the period discussed in Figure 2 where bacterial cultures started to decline at different times. Later on, the growth rate is relatively slow because of the decrease in nutrient concentration and increase in bacterial concentration. However, the final concentrations of bacterial cells are very close, which depends on the amount of broth. This indicates that the maximum concentration in the stationary phase is totally dependent on nutrient concentration, as expected.

It was mentioned how in Figure 1 that we could observe how the absorbance value using light with a wavelength of 630 nm is considerably lower than the absorbance value using a wavelength of 492 nm. It was explained by considering that it is possible for longer wavelengths to miss sufficiently small bacteria, and even the sweetener molecules or ions produced in solution. In order to quantify this loss in detection, we can find a percent difference as shown in Table 2.

**Table 2.** Percent difference between 492 nm and 630 nm wavelength absorbances for initial and final values.

| **Sweetener Type** | **Initial Abs. (492nm)** | **Initial Abs. (630nm)** | **Final Abs. (492nm)** | **Final Abs. (630nm)** | **Difference (Initial)** | **Difference (Final)** | **Percent Difference (Initial)** | **Percent Difference (Final)** |
|---|---|---|---|---|---|---|---|---|
| No Sweetener | 0.09891 | 0.05495 | 1.52252 | 1.24081 | 0.04395 | 0.28171 | 44.44% | 18.50% |
| Fructose | 0.12083 | 0.05963 | 1.17293 | 0.93513 | 0.06119 | 0.23780 | 50.64% | 20.27% |
| Glucose | 0.12994 | 0.06345 | 1.26978 | 0.95251 | 0.06649 | 0.31727 | 51.17% | 24.99% |
| Lactose | 0.10960 | 0.06479 | 1.40425 | 1.04131 | 0.04481 | 0.36294 | 40.88% | 25.85% |
| Maltose | 0.12396 | 0.06521 | 1.16047 | 0.83353 | 0.05875 | 0.32694 | 47.39% | 28.17% |
| Sucrose | 0.12022 | 0.05583 | 0.80610 | 0.61245 | 0.06439 | 0.19365 | 53.56% | 24.02% |
| Glycerol | 0.10125 | 0.05559 | 1.47565 | 1.10519 | 0.04566 | 0.37046 | 45.10% | 25.11% |
| Sorbitol | 0.12170 | 0.05985 | 1.15704 | 0.84133 | 0.06185 | 0.31571 | 50.82% | 27.29% |
| Blue Agave | 0.12339 | 0.06224 | 1.18784 | 0.92609 | 0.06115 | 0.26175 | 49.56% | 22.04% |
| Table Sugar | 0.12875 | 0.05977 | 1.28814 | 0.97677 | 0.06898 | 0.31138 | 53.58% | 24.17% |

Looking at the percent differences, we can observe that the final percent differences are significantly smaller than the initial percent differences for all samples. If we relate the percent difference to the relative concentrations of bacteria and nutrients, the nutrients were at their highest concentration in the LB broth when bacteria were at their lowest concentration. Therefore, the initial measurements might be the most impacted by optical properties caused by nutrients in the broth rather than being impacted by the bacteria. Upon reaching the later stages of growth, the concentration of bacteria has significantly increased while the nutrient density has decreased. This would imply that the concentration of bacteria would then become the dominant factor by the end, while the nutrients reduce in importance.

To address this hypothesis, the absorbance of LB broth without inoculated bacteria was also tested to check for inherent differences in absorbance and is displayed in Table 3. The percentage differences were calculated by considering the average absorbance of the LB broth without sweeteners as a standard. The formula was therefore $\frac{|A_1 - A_X|}{A_1} \times 100$, where $A_1$ is the average absorbance of the LB broth without sweeteners at a given wavelength and $A_x$ is the average absorbance of the sample being compared at the same wavelength.

**Table 3.** Average absorbance, standard deviation, and percentage difference of LB broth with and without sweeteners.

| Sweetener Type | Average Absorbance (492 nm) | Standard Deviation (492 nm) | Percentage Difference with No Sweetener (492 nm) | Average Absorbance (630 nm) | Standard Deviation (630 nm) | Percentage Difference with No Sweetener (630 nm) | Difference Between 492 nm and 630 nm |
|---|---|---|---|---|---|---|---|
| No Sweetener | 0.11060 | 0.00720 | 0.00% | 0.05940 | 0.00700 | 0.00% | 0.05120 |
| Fructose | 0.12527 | 0.00132 | 13.26% | 0.06100 | 0.00065 | 2.69% | 0.06427 |
| Glucose | 0.13040 | 0.00857 | 17.90% | 0.06267 | 0.00413 | 5.50% | 0.06773 |
| Lactose | 0.11171 | 0.00547 | 1.02% | 0.05992 | 0.00530 | 0.91% | 0.05179 |
| Maltose | 0.14560 | 0.00130 | 31.65% | 0.08193 | 0.00096 | 37.93% | 0.06367 |
| Sucrose | 0.12060 | 0.00270 | 9.04% | 0.05737 | 0.00069 | 3.42% | 0.06323 |
| Glycerol | 0.11563 | 0.01050 | 4.56% | 0.06479 | 0.01187 | 9.12% | 0.05083 |
| Sorbitol | 0.14233 | 0.00237 | 28.69% | 0.08250 | 0.00135 | 38.89% | 0.05983 |
| Blue Agave | 0.14077 | 0.00136 | 27.28% | 0.07723 | 0.00026 | 30.02% | 0.06353 |
| Table Sugar | 0.12243 | 0.00193 | 10.70% | 0.05853 | 0.00053 | 1.46% | 0.06390 |

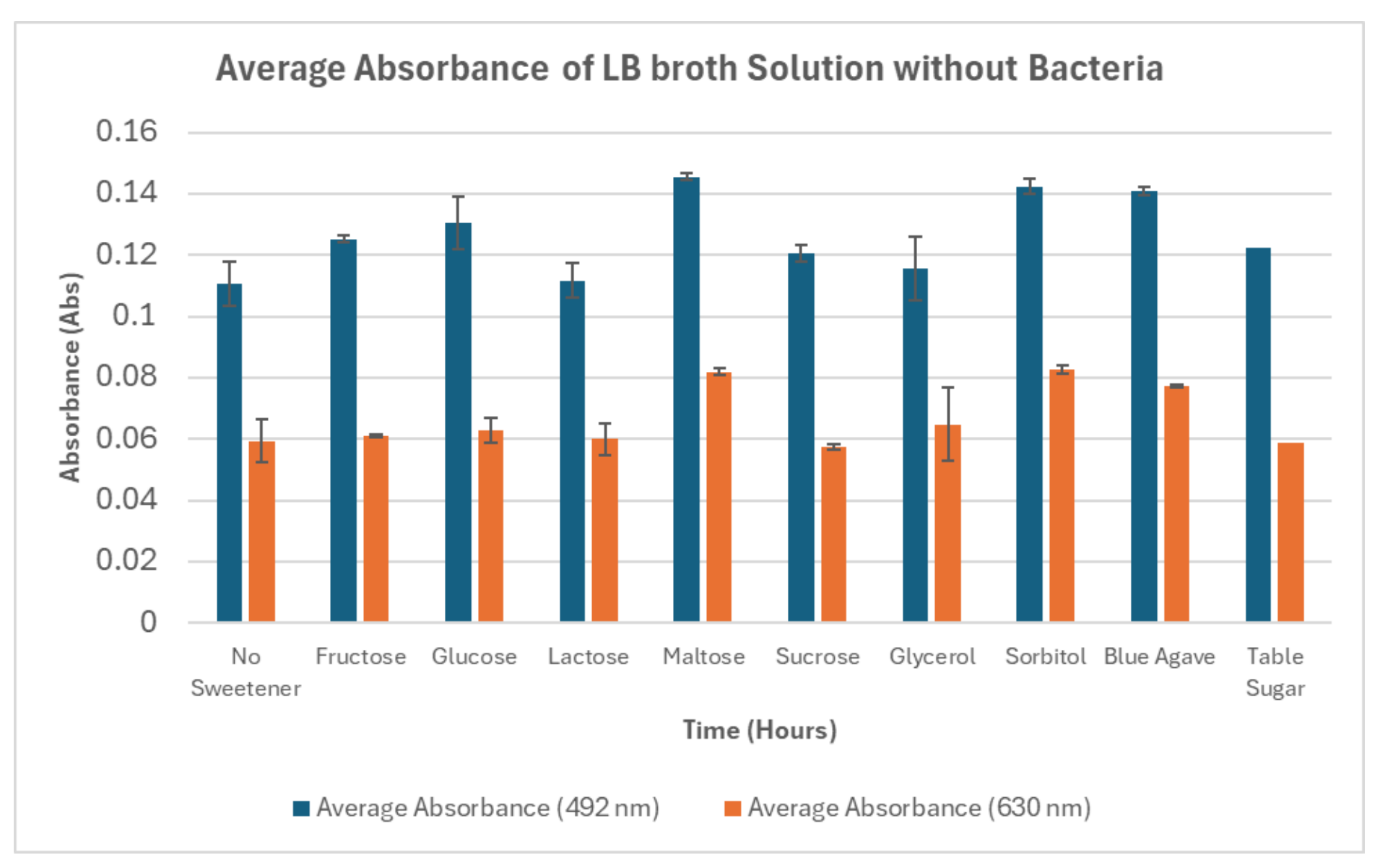

**Figure 4.** Average absorbance values of plain LB broth solutions with and without sweeteners ± 1 standard deviation.

From these results, the difference in absorbance of the LB broth mixtures compared to no sweeteners was largest for maltose with a 31.65% difference compared to pure LB broth. This implies that maltose is expected to have a significant absorbance contribution. However, the largest difference in absorbance values when varying the measuring wavelength is found in the LB broth containing glucose. This difference is 0.06773, and each of the other differences due to the wavelength are around 0.05 or 0.06. These values are comparable to the initial differences measured in Table 2. Therefore, the initial percent differences in Table 2 are mostly due to the LB broth solution. However, the same doesn't appear to be true for final absorbances. The final differences in Table 2 lie between 0.19 and 0.37, which are both many times greater than the difference due to varying the measuring wavelength. Therefore, the nutrient composition is a smaller factor by the end and the concentration of bacteria is the primary factor for the difference in absorbance seen in Table 2. This means that 492 nm is likely more sensitive for detecting these bacteria.

Previously, it was noticed that, after 10-12 days of growth at room temperature, each of the samples containing sweeteners reached higher pH values. However, after 5 days of growth within the magnetic field environments at room temperature, pH values were estimated as can be seen in Table 4. Samples containing sweeteners tended to receive lower pH estimates, with the sugar samples (both lab grade and consumer grade) being the most acidic (lowest pH) and sugar alcohol samples being only slightly acidic. Averages of each set of 3 samples can also be seen in Figure 5.

**Table 4.** Estimates of pH after 5 days of growth in various magnetic field environments.

| **Magnetic Field Type** | **B = 0** | | | **B ≡ RA** | | | **B ≡ OS** | | | **B ≡ STA** | | |
|---|---|---|---|---|---|---|---|---|---|---|---|---|
| **Sample Number** | **#1** | **#2** | **#3** | **#4** | **#5** | **#6** | **#7** | **#8** | **#9** | **#10** | **#11** | **#12** |
| **No Sweeteners, pH** | 7 | 7 | 6.5 | 6.5 | 6 | 6 | 6.5 | 7 | 7 | 7 | 7 | 7 |
| **Fructose, pH** | 4 | 4 | 4 | 3.5 | 3.5 | 3.5 | 4 | 4 | 4 | 4 | 3.5 | 4 |
| **Glucose, pH** | 4 | 4 | 4 | 4 | 3.5 | 4 | 4 | 3.5 | 3.5 | 3.5 | 3.5 | 4 |
| **Lactose, pH** | 3.5 | 3.5 | 4 | 4 | 4 | 4 | 4 | 3.5 | 4 | 4 | 3.5 | 3.5 |
| **Maltose, pH** | 5 | 5 | 5 | 4 | 4 | 4 | 4 | 4 | 4 | 4 | 4 | 4 |
| **Sucrose, pH** | 5 | 5 | 4 | 3.5 | 4 | 4 | 4 | 4 | 4 | 4 | 4 | 4 |
| **Glycerol, pH** | 6 | 5 | 5.5 | 5 | 5.5 | 5.5 | 6 | 6 | 5.5 | 5.5 | 5.5 | 5 |
| **Sorbitol, pH** | 6 | 6 | 6 | 6 | 6 | 6 | 6.5 | 6.5 | 6.5 | 6.5 | 6.5 | 6.5 |
| **Blue Agave, pH** | 5 | 4 | 4 | 4 | 4 | 4 | 4 | 4 | 4 | 4 | 4 | 4 |
| **Table Sugar, pH** | 4 | 4 | 4 | 4 | 4 | 4 | 4 | 3.5 | 4 | 3.5 | 3.5 | 4 |

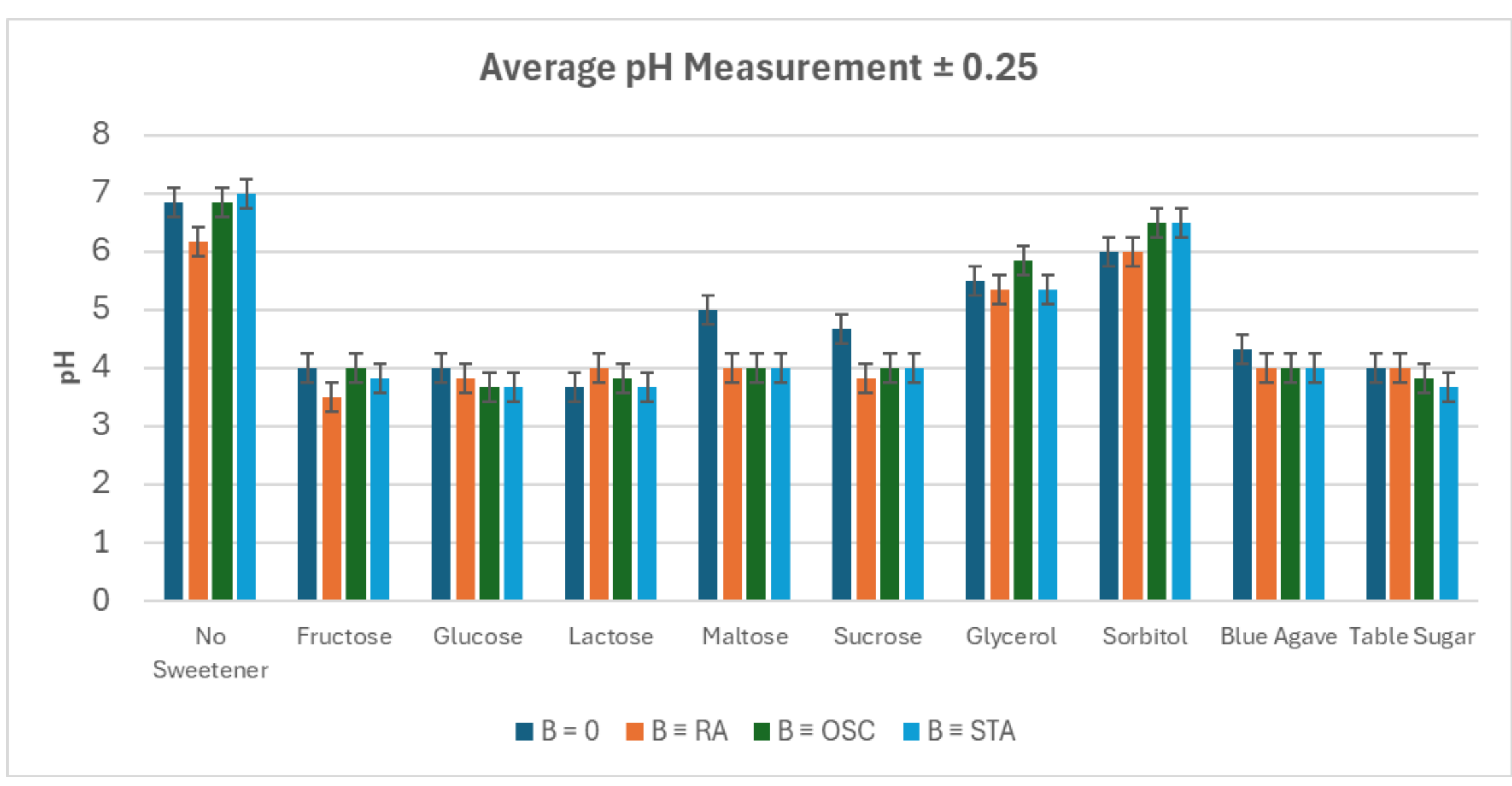

**Figure 5.** Averages of pH estimates from Table 4, with error bars of plus or minus 0.25.

Results of pH values correspond to estimated values (Table 4), as test strips were compared with a manufacturer-provided color scale to determine the values. As expected, the estimated pH of the sugar alcohol samples did not differ from the sweetener-free samples to the same degree as the sugar samples. This may be due to byproducts produced from the metabolism of sweeteners by the bacteria, causing a change in the pH of the solution. The reduction in pH for the samples may be a contributing factor causing the relative inhibited growth of the samples, with the greatest pH reduction from sugars causing the greatest inhibiting effect. The most notable differences regarding magnetic field configurations are shown in no sweetener samples with the B ≡ RA magnetic field samples having slightly lower pH, and with maltose and sucrose samples which have a considerably higher pH in B = 0 samples. These differences may indicate minor chemical changes caused by perturbations originating from the magnetic fields which can affect the rate of bacterial growth [4].

## 3. Materials and Methods

### 3.A Materials

A mother culture was prepared in nutrient broth (Luria-Bertani broth or LB broth) for this experiment. A bacteria colony was collected from an initially prepared sample using standard techniques of plating and the selection of an individual colony from a batch of *S. aureus* identified as NCIMB 12702.

An initial mixture of LB broth (Miller's) was prepared by adding premixed powder to purified water with the ratio 25 g/L and autoclaving at 121 °C and a pressure of 1.2 bar (17.4 psi) inside a Fisher Scientific SterilElite 16 SA 230FA autoclave. After autoclaving was complete, the LB broth was allowed to momentarily cool to a more safely manageable temperature. Experimental samples were then prepared by filling sterilized test tubes with approximately 5 mL of LB broth. 12 samples were made for each sweetener configuration. For each sweetener sample, 40 μL was added to each tube of an aqueous solution with the exception of glycerol samples, where 20 μL of non-aqueous glycerol was added to each tube. The "no sweetener" samples contained LB broth with no additives prior to inoculation.

The brand of blue agave syrup used was Kirkland Signature Organic Blue Agave, while the consumer-grade table sugar was granulated sugar from Smart Way. The authors have no relationship with these brands, and these products are listed only for informational purposes.

Tests were done using a microplate and a plate reader to study growth rates based on measurements of absorbance. We employed a Corning Inc. 96 well culture plate and an Accuris® Smartreader 96 microplate absorbance reader for this purpose. We had previously used a Thermo Scientific spectrophotometer (Spectronic 20D+), where each individual tube's optical density (OD) was measured individually. However, although the results obtained by both methods are comparable, the Spectronic 20D+ presented an increased chance of error due to the required manual measurement of each individual tube as compared to plate reader which measured every sample almost simultaneously according to a schedule. In addition, we took all measurements at 580 nm and these results may differ for other wavelengths, as demonstrated in our results. The use of a scheduled automatic device also reduces human fatigue, resulting in further reduction of possible error sources.

For pH tests, Hydrion pH strips were used which show the pH from 1.0 to 12 based on the resulting color of the strip after contact with the sample.

### 3.B Experimental Setup

Four environments with differing electromagnetic fields were prepared, 3 with different magnetic fields and one with no magnetic field as a control. Each environment was located within the same lab at temperatures of approximately 21.5 °C. We left our control samples at a sufficiently large distance from the magnets to keep them away from the magnetic field effects and considered it as the first environment as the magnetic field measurement was set to zero to calibrate the Gauss meter.

The second magnetic field environment was referred to as the "random" magnetic field ($B \equiv RA$), shown in Figure 6a. This field was generated by a magnetic strip coiled into a thin, wide disk. The magnitude of the magnetic field close to the strip lies within ±2.0 mT, and the field changes rapidly with miniscule changes in position above the strip. This results in an unpredictable and highly position-sensitive field gradient, identified as the "random" field.

The third environment was an oscillating field ($B \equiv OS$), as shown in Figure 6b. This field was generated using a function generator and 10 sideways and stacked 200-turn coils connected in series as shown in Figure 6c. The stacked and interconnected coils acted as a solenoid, while the function generator controlled the voltage. The function generator was set to generate a sinusoidal wave with a voltage of 7.50 V and a small frequency of 0.5 Hz. The measured magnitude of the resulting sinusoidal wave was approximately ±1.0 mT, as can be seen in Figure 12b. No measurable change in temperature within the solenoid was observed after the system remained operational for approximately 5 days.

The fourth magnetic field environment was a static field environment ($B \equiv STA$) generated by a constant current flowing through a stack of interconnected 200-turn coils, as shown in Figure 6d. The setup for this environment was almost identical to the $B \equiv OS$ environment, except a DC power source with constant voltage was used instead of a function generator. The voltage was set to 13.5 V, resulting in a current of 0.25 A and a field magnitude of approximately 2.0 mT (or 20 Gauss). As with the oscillating field, no measurable change in temperature within the solenoid was observed after the system remained operational for approximately 5 days.

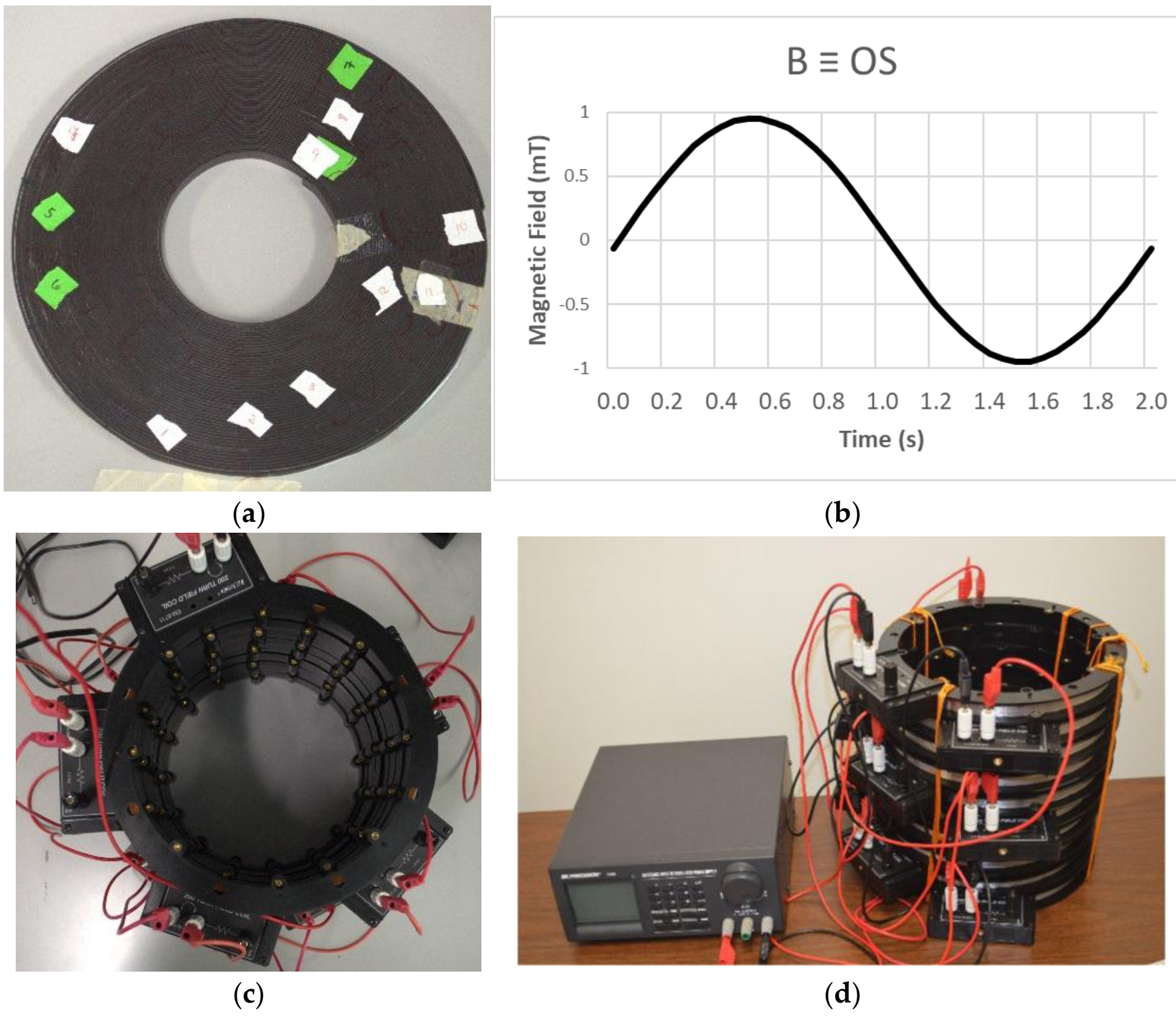

**Figure 6.** (**a**): Magnetic field environment created by winding a magnetic strip. Referred to as the random magnetic field due to position sensitivity. (**b**): B ≡ OS (oscillating magnetic field) field magnitude over time. (**c**): Magnetic field environment created by a combination of stacked Helmholtz coils connected in series, resulting in a constant transverse field with a magnitude of around 2.0 mT (20 Gauss). (**d**): shows a simple arrangement of Helmholtz coils to which we provide a small current of 10 Amp to produce a constant transverse magnetic field of 20 Gauss.

This experiment focused on the growth of *S. aureus* with sweeteners, which were compared with a control group containing no sweeteners. The experiment was conducted three times with the same procedure for each run. The first time, samples with no sweetener, lactose, and glycerol were tested. The second run used sucrose, glucose, fructose, and table sugar, and the third test used sorbitol, maltose, and blue agave. Samples were prepared in test tubes with a combination of LB broth, sweeteners, and bacteria samples from the mother culture. An LB broth solution was made from which 5 mL was placed in each test tube, and as for the sweeteners, the desired concentration was 5 g/L (0.025 g for 5 mL). Considering that the glycerol was in liquid form having a density of 1.26 g/mL, the amount that was added for the glycerol samples was 20 µL. For the rest of the sweeteners, they were diluted in water to obtain aqueous solutions where 40 µL contained 0.025 g of the sweetener and 40 µL of each solution was then added to each respective sample. All sweetener solutions were briefly heat-treated at approximately 100 °C before use to ensure sterility, and each sweetener had 12 sample test tubes for the experiment.

Each plate reader microplate contains 96 wells, and when testing, each sweetener type had a dedicated row containing an LB broth solution with *S. aureus* and a specific sweetener. For each row, 12 wells with the same sweetener solution (or no sweetener) were filled. Since not all sweeteners were tested at the same time, not all wells were filled for each test. Other microplates were later filled with sterilized LB broth (no bacteria) with and without the same added sweeteners as a control to measure the absorbance of the LB broth solutions without bacteria. For samples containing bacteria, the wells were filled with a sample of 20 μL of the respective LB broth with bacteria and an additional 180 μL of sterile LB broth solution, bringing them up to 200 μL of solution per well. The control LB broth wells were filled with 200 μL of their respective sterilized LB broth solutions.

Each of the test tube samples were grown in one of the four prepared electromagnetic field environments, splitting each sweetener set of 12 tubes evenly between the environments. This resulted in 3 tubes per sweetener being used for each of the environments. For both the static and oscillating fields, the tubes were grouped together held upright inside beakers which were placed in the middle of the solenoids. The beakers were also raised by placing plastic test tube holders beneath them such that the tubes would be centered both radially and vertically within the solenoids. For the random field, the tubes were placed horizontally over the wound magnetic strip in marked positions to standardize positions and to minimize the distance between the samples within the tubes and the magnetic strip. For the samples placed away from magnetic fields, samples were placed in a test tube rack far away from the other magnetic field setups such that no significant magnetic field was measured. While the test tube cultures were being grown in these environments, the plate reader samples were also grown separately in the microplate wells and analyzed using the plate reader.

Samples were left to grow for 5 days in their respective environments. For the samples in the plate reader, the device was set up to take absorption measurements every 4 hours for the same 5 days. Absorbance measurements were completed using wavelengths of 492 nm and 630 nm, with gentle shaking for approximately 30 seconds before each reading. For pH measurements, these were taken for each of the tubes using pH strips. Since this process involves taking out liquid samples from inside the test tubes, to minimize the risk of contamination, pH measurements were only taken once after the 5-day period had passed. Each tube was vortexed prior to pipetting small volumes of sample onto each test strip for analysis, and pH readings were estimated within increments of 0.5 according to the color scale provided by the manufacturer.

## 4. Conclusions

It can be seen as a general trend that the maximum bacterial growth in any type of sweetener is less than the growth in pure broth. However, where no sweetener samples reach a stationary phase after 72 hours of growth, other samples including sucrose, glucose, and table sugar trend upwards towards the end of the 5 days pointing at the possibility of further long-term growth.

Our results indicate that adding both lab grade and consumer grade sugars to a solution of LB broth and *S. aureus* causes the pH of the solution to decrease as bacteria reproduce, which may inhibit further bacterial growth. This effect can be attributed to byproducts of sugar metabolism increasing acidity. Adding sugar alcohols to the solution appeared to have similar effects on the pH, but to a lesser extent, and as result, growth appears to be not as inhibited as glucose, fructose, and their derivatives (sucrose, table sugar, blue agave). This pattern is, however, not consistent with maltose samples, made of two glucose molecules, and lactose samples, which contain a galactose molecule aside from glucose.

The standard deviation of absorbance measurements was greatest in the late stages of growth for samples such as sorbitol due to different cultures entering death phase at varying times, and a similar effect can be observed for other sweetener types but to a lesser extent. However, measurements before this stage seem to be consistent among samples, producing a reasonable standard deviation during early growth. Based on this data, the greatest growth inhibition and greatest acidity both seem to correlate to the addition of normal sugars within samples, while lesser inhibition and acidity correlate with the addition of sugar alcohols within samples. That said, this pattern seems to differ with lactose and maltose samples, which might be linked to differences in composition.

pH measurements show slight changes based on magnetic field environment. Specifically, the "no sweetener" samples show a decrease in pH for $B \equiv RA$ samples, and maltose and sucrose show an increase for $B = 0$ samples. While this difference is noticeable, other sugars do not show any noticeable patterns or their differences fall under expected error. As a result, further studies might be required to study magnetic field effects by either analyzing growth under magnetic fields, which would require different experimental set ups, or by exploring magnetic fields with varying magnitudes.

The composition of LB broth with or without sweeteners was the primary factor affecting the absorbance of samples initially, but at later stages, the composition became a less significant factor affecting the absorbance. The choice between wavelengths of 492 nm and 630 nm significantly altered the absorbance measurements, especially for readings with larger absorbance values, implying that the shorter 492 nm was more sensitive to detecting the small bacteria cells of *S. aureus*. Differences in peak absorbance values for the sweeteners may also indicate that the bacteria tend to grow to different sizes before cell division. Further study of other sugars or sugar alcohols as additives to solutions of LB broth and *S. aureus* are necessary to verify potential effects on the bacteria size.

Because bacteria morphology can play a role in the infectious behavior of bacteria, morphological changes caused by varying bacterial size due to the presence of sweeteners may be of further interest. Additionally, there are some indications of modifications in the interaction of antibiotics with bacteria in the presence of magnetic fields [17-18]. For this reason, an extensive study is needed in this direction looking at any morphological changes that may be caused by the presence of sweeteners or magnetic fields.

It is important to mention that exposure to weak fields in this era of technological development is almost unavoidable. Similarly, sugars and sugar alcohols can be commonly found in our food. The combined impact of sugars, sugar alcohols, and magnetic fields on cellular mechanisms and growth rates can help a lot to protect human health and our environment on this planet. In particular, weak magnetic fields may be of significant interest in environments where other fields such as gravity are insignificant yet electronics are plentiful, such as in spacecraft. Therefore, understanding how weak magnetic fields interact with microbes such as those found in the human microbiome is an important topic of space biology. *S. aureus* is one of many such microbes commonly found in the human microbiome [19], so further study is necessary for a complete picture of human health during space travel.

**Author Contributions:** Conceptualization, Samina Masood; Methodology, Samina Masood; Investigation, Angel Arrieta and Derek Smith; Data curation, Angel Arrieta and Derek Smith; Writing – original draft, Samina Masood; Writing – review & editing, Angel Arrieta and Derek Smith; Supervision, Samina Masood. All authors have read and agreed to the published version of the manuscript.

**Funding:** This research received no external funding.

**Institutional Review Board Statement:** Not applicable.

**Informed Consent Statement:** Not applicable.

**Data Availability Statement:** The data that support the findings will be available in BIOPHYSICS LAB at https://www.uhcl.edu/science-engineering/faculty/masood-samina

following an embargo from the date of publication to allow for commercialization of research findings.

**Conflicts of Interest:** The authors declare no conflict of interest.